\documentstyle[12pt,aaspp4,psfig]{article}

\def\lesssim{\mathrel{\hbox{\rlap{\hbox{\lower4pt\hbox{$\sim$}}}\hbox{$<$}}}}
\def\gtrsim{\mathrel{\hbox{\rlap{\hbox{\lower4pt\hbox{$\sim$}}}\hbox{$>$}}}}

\begin{document}
\pagestyle{myheadings}

\markright{Fruchter, Krolik, \& Rhoads 2001 ---
Accepted to {\it The Astrophysical Journal} \hfill}

\def\apjs{Ap. J. Suppl.}

\title{X-ray Destruction of Dust Along the Line of Sight to $\gamma$-ray 
Bursts}

\author{Andrew S. Fruchter}
\affil{Space Telescope Science Institute, Baltimore, MD 21218}    

\author{Julian H. Krolik}
\affil{Department of Physics and Astronomy, Johns Hopkins University, 
	   Baltimore, MD 21218}
\and
\author{James E. Rhoads}
\affil{Space Telescope Science Institute, Baltimore, MD 21218}    

\begin{abstract}

     We show that if all $\gamma$-ray bursts emit X-rays in a way similar to
those observed by BeppoSax, much of the extinction along the line of sight
in the host galaxy of the burst can be destroyed.  Two mechanisms are
principally responsible for dust destruction:
grain heating and grain charging.  The latter, which can lead to electrostatic
stresses greater than the tensile strength of the grains, is often the more
important.   
Grains may regularly be destroyed at distances
as large as $\sim 100$~pc.  This dust destruction
can permit us to see the UV/optical afterglow even when the burst is embedded
deep within a highly-obscured region.
Because the destruction rate depends on grain composition and
size, it may be possible to observe the amount and wavelength-dependence of
extinction change during the course of the burst and first few minutes
of the afterglow.
It may also be possible to detect interstellar absorption lines in the
afterglow spectrum that would not exist but for the return of heavy
elements to the gas phase.

\end{abstract}

\section{Introduction}

Afterglows currently  afford the best available diagnostics of 
$\gamma$-ray burst environments.  Although so far
only afterglows from long, soft bursts -- a subclass containing about
two-thirds of GRBs (Kouveliotou et al. 1993) -- have been localized,
these localizations have resulted in 
redshift measurements
of the bursts (e.g., Metzger et al. 1997), as well as
(subarcsecond) burst positions, 
and identification and studies of the host galaxies
(Hogg \& Fruchter 1998;
Mao \& Mo 1998; Fruchter et al. 1999;  Bloom et al. 1999).
One of the most revealing results from
afterglow studies is a range of observations possibly suggesting that 
$\gamma$-ray
bursts are associated with massive star formation and the associated
regions of dense interstellar gas:

(1) Host galaxies of $\gamma$-ray bursts show systematically blue
broadband colors (Fruchter et al. 1999).

(2) A number of host galaxies show unusually strong [\ion {O}{2}]
emission (Bloom et al. 1998, Djorgovski et al. 1998,  Vreeswijk et al. 2000).

(3) The wide range of optical to $\gamma$-ray flux ratios may be explained
if there is substantial extinction along the line of sight to some
bursts.  The best documented case here is GRB 970828 (Groot et al.
1997), for which an extreme $\gamma$-ray to optical ratio was combined
with evidence of soft X-ray absorption, suggesting a large column
density of ISM at a substantial redshift.

4) Features in the light curves and color evolution of GRBs 980326 and
970228 have been interpreted at least by some 
(Bloom et al. 1999, Galama et al. 2000)
as evidence for associated supernovae, again suggesting a link between
GRBs and the deaths of stars sufficiently massive to remain near their
natal star forming regions throughout their lives.

(5) The peculiar spectral energy distribution of the GRB 980329
afterglow (Fruchter 1999, Reichart et al 1999) may be due to 
absorption bands of excited molecular hydrogen in the vicinity
of the burster (Draine 1999).  This would require a considerable
column density of molecular hydrogen  ($\sim 10^{18} {\rm cm}^{-2}$)  
in the neighborhood of the burster.

6)  Although the measured extinction of the optical transient associated
with GRB~980703 is not particularly high ($A_V = 1.5 \pm 0.11$~mag), 
the low-energy X-ray spectrum of the afterglow
is best fit by  $N_{\rm H} = 3.6^{+2.2}_{-1.3} \times 10^{22}$~cm$^{-3}$,
suggesting that the burst may nonetheless have occurred in 
a molecular cloud (Vreeswijk et al. 1999).  Other bursts may be similar
to GRB~980703 in this regard (Galama \& Wijers 2000).

All of these lines of evidence, then, point to a possible
association of at least the long duration, soft spectrum
bursts with the dense ISM, and hence with potentially high dust column
densities.  If large dust column densities are common, then a large
fraction of afterglows could be obscured at optical through soft X-ray
wavelengths.  However, many afterglows-- a majority of those observed
at optical wavelengths-- are in fact relatively blue in their optical colors.

The association of GRBs with star formation may be best reconciled
with observed blue colors if GRBs destroy much of the dust
along the line of sight to the burst.  GRBs are extremely energetic
events at all wavelengths, making many dust
destruction mechanisms possible.  Waxman and Draine (2000) were the first to
explore some of these mechanisms in detail.  They examined
particularly the evaporation of dust grains through direct heating by 
UV radiation, and (more briefly) destruction through photoelectric
grain charging.  Using a reverse shock model, 
they
scaled the prompt UV/optical emission of GRBs
from the single detection of prompt optical emission to date (GRB
990123; Akerlof et al 1999), which now appears to have been
exceptional in its optical and ultraviolet emission (e.g., Williams et al
1999, Akerlof et al. 2000).

In the present paper, we examine grain evaporation through heating by
X-rays, which can dominate over the heating by UV photons, 
and also give a more detailed treatment of grain shattering
by photoelectric charging.  Our closer examination of the electrostatic
grain shattering mechanism leads to a significant increase in its relative
importance.  Our results imply that the amount of dust destroyed by X-rays
associated with GRBs may considerably exceed that destroyed by the prompt
UV pulse alone.

Although there is a great degree of uncertainty regarding possible
collimation and beaming of $\gamma$-ray bursts, our results are very
nearly independent of these questions.   Only the X-rays radiated along
our line of sight and the dust grains lying along that same line of sight
are relevant to the issues we pursue.  So long as the photons in the
optical/UV afterglow travel within the cone filled by the X-ray beam,
it does not matter how large those solid angles are.

\section{Phenomenology of X-ray and Optical Emission Associated with
$\gamma$-ray Bursts}

When considering destruction of dust, we are primarily concerned with the
spectral window from the X-ray through the optical, where the cross
section for photons to interact with dust grains is largest.  Dust grains
can absorb optical and ultraviolet photons through a variety of continuous
opacity mechanisms; X-rays are primarily absorbed by K-shell ionization
of medium-Z elements such as C, Si, O, etc.  The X-rays with which we
are primarily concerned are therefore those whose rest-frame energies
lie in the band from 0.3 to $\simeq 10$~keV,
a stretch defined by the C K-edge at the low-energy end and the Fe
K-edge (plus a small energy margin; see below) at the high-energy end.

  In the case of the long (1 -- 100~s) duration bursts, 
the X-ray emission is now known to be broken into two
parts: a short (1 -- 100~s) pulse, largely coinciding with the
$\gamma$-ray burst itself; and a longer, smooth decline,
known as the afterglow.  The short duration bursts ($<1$s)
have been inaccessible to BeppoSax, and their afterglow properties,
if any, are at present unknown.  

During the burst proper, the X-ray flux in the 0.3 to 10~keV band is
strong and hard, with an energy spectral index $\alpha$ (flux
per unit energy $\propto \epsilon^{-\alpha}$ for photon energy $\epsilon$)
typically between -1 and 1 (Band et al. 1993, Frontera et al. 2000). 
During the peak of the burst, this hard spectrum can continue up
to or above 100~keV before it breaks.   However, as the burst progresses,
the break drops to lower energies and the X-ray spectrum softens, leading
to a temporary increase in the 0.3 -- 10~keV band (Frontera et
al. 1999).  As the burst fades, a prolonged afterglow
phase begins.

  Although the physics which produces the rapid, variable $\gamma$-ray and 
X-ray burst is not particularily well understood, there is 
good evidence that the afterglow is produced by synchrotron
emission from electrons accelerated in the external shock between GRB
ejecta and an ambient medium as predicted 
by Paczy\'{n}ski \& Rhoads (1993), Katz (1994) and
Meszaros \& Rees (1997).  This class of model leads to a broken
power law spectrum whose break frequencies evolve as power-laws in
time (Sari, Piran, \& Narayan 1998), and has been reasonably
successful at describing the observations.

Most afterglows are well fit by power law slopes $p \approx 2.3$ for
the injected electron energy spectrum in the expanding fireball's
external shock.  The highest energy spectral break in the afterglow
spectrum is usually the cooling break.  At frequencies above the
cooling break, the temporal decay of afterglow emission goes as
$t^{1/2 - 3p/4}$, while at lower frequencies, it follows a shallower decay of
$t^{3/4 - 3p/4}$.  For observed afterglows, the cooling break passes through
the X-ray band very early (observed $t\ll 1$ day), but its passage
through the optical band can occur anywhere from $t\ll 1$ day to $t
\sim 2$ days (Galama et al 1999).  This means
that the X-ray afterglow often fades faster than the optical
afterglow.

As we will show below, both the X-ray flux and fluence are important in dust
destruction by GRBs.  The peak X-ray flux occurs during the GRB
itself, with characteristic values $\sim 10^{-7}$~erg~cm$^{-2}$~s$^{-1}$
for bursts detected by BeppoSAX.  The
relative contributions of the afterglow and the GRB itself to the
X-ray fluence are discussed by Frontera et al (2000), who find that
each contributes $\sim 10^{-6}$~erg~cm$^{-2}$ to the
X-ray fluence from BeppoSax bursts, although in individual cases
the ratio of burst to afterglow X-ray fluence has been anywhere from
$\sim 1/3$ to $\sim 3$.  (Practically, the fluence during the burst is
determined by direct integration of the observed light curve during
the duration of measurable gamma-ray emission, while the fluence of
the afterglow is determined by fitting power law light curves to the
observed points on the afterglow light curve.  This phenomenological
division is relevant for our calculations whether the physical
mechanisms responsible for X-ray emission during the GRB and the
afterglow differ or not.)  The X-ray emission during the GRB is
generally consistent with the extrapolation of the X-ray afterglow to
early time (Frontera et al 2000).
For typical afterglow decay rates, at least half the X-ray afterglow
fluence is typically received within $10
\times \Delta t_{GRB}$, where $\Delta t_{GRB}$ is the duration of the
GRB (see Dal Fiume et al 2000 for further discussion of this point).

    Given how rapidly even the afterglow X-ray flux decays, we conclude
that any effects on intervening dust due to X-ray irradiation take
place within a few minutes of the burst.  Thus, any attempts to
watch dust destruction in progress will require very fast response.

\section{Effect of X-rays on Dust}

\subsection{Elementary interactions}

      The fundamental interaction between X-rays and dust grains is
K-shell photoionization of medium-$Z$ elements such as O, Si, S, and Mg,
and either K-shell or L-shell photoionization of Fe.  
When K-shell photoionization occurs, a
fast electron is created with energy equal to the difference between the
absorbed photon energy and the ionization threshold for the relevant
shell.  Very soon after, the atom relaxes by one of two processes:
Auger ionization (which results in a second fast electron with energy
slightly less than the initial ionization threshold) or fluorescence.
For elements in this range of atomic number, Auger ionization usually
dominates.

      As these fast electrons travel through the grain, they lose
energy by Coulomb scattering other electrons.  In some cases
the scattered electrons also begin travelling through the grain.  However, no
electron gets very far unless its initial energy is fairly high.  According
to Draine \& Salpeter (1979), the range of a fast electron (energy $300$ 
eV to $1$ MeV) in typical interstellar dust materials is
\begin{equation}
R \simeq 0.03 \rho^{-0.85} E_{\rm keV}^{1.5}\, \mu\hbox{m},
\label{range}
\end{equation}
where $\rho$ is the density in gm~cm$^{-3}$ and $E_{\rm keV}$ is the initial
electron energy in keV.  Thus, in grains as large as $\sim 0.1$~$\mu m$,
only electrons either more energetic than several keV, or produced
very close to the grain surface, can escape (see Dwek \& Smith 1996 for
a more detailed treatment of the effects of photoionization in dust
grains).  Below we will discuss a
second mechanism that can also inhibit electron loss.

     Electron Coulomb scattering also has another effect on the grain. 
Significant energy transfer to secondary electrons is likely to
break chemical bonds, weakening the crystal structure.  We will return
to the consequences of this fact later.

\subsection{Heating and sublimation}

      The immediate result, however, of many X-ray photoionizations is 
simply the heating of an individual grain at a rate of
\begin{equation}
G \approx 3 \times 10^{-3} {{\cal T}_h (x_K,\alpha,N) \over 2+\alpha}
{E_{51} \sigma_{-19} n_{23} \over D_{100}^2} 
{a_{0.1}^3 \over t_{10}} x_K^{1-\alpha}
\left[1 - \left(1 + x_{min}/x_K\right)^{-(2+\alpha)}\right]
\hbox{~erg s$^{-1}$},
\label{heating}
\end{equation}
where the grain is assumed to be
optically thin to the X-rays, as is generally the
case.  Here $E_{51}$ is the total energy radiated in X-rays in the
sense of $4\pi\epsilon dE_\epsilon /d\Omega$, evaluated at
$\epsilon_o = 1$~keV and scaled to $10^{51}$~erg, and
$dE_\epsilon/d\Omega$ is the energy radiated per unit energy per solid
angle.  Other symbols are:
$\sigma_{-19}$, the K-shell edge cross section in units of $10^{-19}$~cm$^2$;
$n_{23}$, the atomic density in the grain in units of $10^{23}$~cm$^{-3}$;
$x$, photon energy in keV;
$a_{0.1}$, the grain size in units of $0.1$~$\mu m$;
$t_{10}$, the characteristic time of the X-ray emission scaled to 10~s;
$D_{100}$, the distance from the $\gamma$-ray burst source to the dust in
units of 100~pc;
$\alpha$, the (energy) spectral index, defined in the sense that
$E_\epsilon \propto \epsilon^{-\alpha}$;
$N$, the H atom column density along the line of sight;
$x_K$, the energy in keV of the most important K-edge in the grain;
and $x_{min}$, the minimum energy (in keV) of electrons able to escape from
the grain (see more extensive discussion below).
In deriving this equation, we have taken $\sigma \propto x^{-3}$ for $x > 
x_K$. 
In practice, as noted before, the low-energy X-ray spectrum of GRBs
during the burst varies both 
between bursts and during individual bursts, with values ranging over the 
interval $-1 \la \alpha \la 1$ (Frontera et al 2000).
The factor ${\cal T}_h$ describes the transparency to X-rays
of the interstellar medium (ISM) between the burst source and the dust grains,
suitably averaged over the X-ray spectrum of the GRB.  We will discuss this
factor at
greater length in \S 4.1.

  It will be convenient to group the combination of fiducial
factors $\sigma_{-19} n_{23} E_{51} D_{100}^{-2}$; we call this quantity the
modified fluence, and symbolize it as $A$.
The combination $\sigma_{-19} n_{23}$ should
be order unity, its exact value depending on the grain composition.  For
example, if the grains are olivine, whose chemical composition is
Mg$_2$ Si O$_4$ (as found in O-rich environments: Waters et al. 1996),
$\sigma_{-19} n_{23} \simeq 2$ for photons above the Si edge at 1.9~keV.
On the other hand, the value of $\sigma_{-19} n_{23}$ for graphite is
$\simeq 17$, with the threshold $x_K = 0.28$.

     Two processes, radiative cooling and sublimation, counterbalance
heating.  Following Spitzer (1978), we write the rate of radiation
cooling per grain as $4 \pi a^2 \times \pi \int Q_a B_\lambda
d\lambda$, where $Q_a$ is the absorption efficiency of the grain and
$B_\lambda$ is the Planck function.  For grains smaller than $\simeq
1$~$\mu$m, $Q_a$ can be considerably less than unity because the
characteristic wavelength of their thermal radiation is rather larger
than their size.  In the small grain limit ($2 \pi a \ll \lambda$),
$Q_a \approx 8 \pi a / \lambda \times \hbox{Im}[(m^2- 1)/(m^2+2)]$, 
with $m$ the complex index of refraction of the grain material and $\hbox{Im}$
denoting the imaginary part.  In the relevant temperature range (i.e.,
$T \sim 2000$~K), for most grain materials $\hbox{Im}[(m^2- 1)/(m^2+2)]$
is almost independent of $\lambda$ over the range of wavelengths that dominate
the integral ($\lambda \sim ch/kT$).  Rather than following the usual
scaling ($\propto T^4$), the cooling rate is then $\propto T^5$.
Typical values of $\hbox{Im}[(m^2- 1)/(m^2+2)]$ are in the range
0.01 -- 0.1; we adopt a fiducial value of 0.065 so that the cooling
rate per grain becomes $3.1 \times 10^{-3}  a_{0.1}^3 T_3^5 \phi
\hbox{~erg s$^{-1}$}$.  With this choice, we can estimate values of
the material-dependent correction factor $\phi$ by comparing to the
cooling rate computed by Waxman \& Draine (2000) for the temperature
range $2000 K \la T \la 3000 K$ based on the optical constants of Draine
\& Lee (1984): $\phi \approx 0.3$ for silicates,
$\phi \approx 3$ for graphite, and $\phi \approx 1$ is a representative
average value.

Following Waxman \& Draine (2000) and Guhathakurta \& Draine (1989),
we write the sublimation rate as
\begin{equation}
{d a \over d t} = -n^{-1/3} \nu_0 \exp\left[-H / (k T)\right]~~.
\label{subl_rate}
\end{equation}
Here the characteristic frequency $\nu_0 \sim 10^{15}$~s$^{-1}$ and binding
energy per atom $H \sim 10^{-11} \hbox{erg}$  depend on the grain material.
Estimates for graphite are $\nu_0 \approx 2 \times 10^{14}$~s$^{-1}$ and
$H \approx 1.1 \times 10^{-11} \hbox{erg}$, while for silicates,
represented by Mg$_2$SiO$_4$, $\nu_0 \approx 2 \times 10^{15}$~s$^{-1}$
and $H \approx 0.94 \times 10^{-11} \hbox{erg}$ (Waxman \& Draine
2000; Guhathakurta \& Draine 1989).  For the purpose of these estimates
(and in the absence of published values for $H$ and $\nu_0$
pertaining to carbon-rich grains with other structures), we will suppose
that the graphite numbers apply to all carbonaceous grains.
Clearly, the sublimation rate is very sensitive
to the temperature.

   Neglecting fluctuations due to absorption of individual
energetic photons (these will be most important in very small grains, of
course), the grain temperature is determined by the heat balance equation
\begin{eqnarray}
\label{heatbal}
{d {\cal Q} \over d t} & = &
3 \times 10^{-3} {{\cal T}_h A \over (2+\alpha) t_{10}} {a_{0.1}^3 }
x_K^{1-\alpha}\left[1 - \left(1 + x_{min}/x_K\right)^{-(2+\alpha)}\right]
-  3.1 \times 10^{-3} a_{0.1}^3 T_{3}^5 \phi \\
& & - 2.7 \times 10^{10} a_{0.1}^2 n_{23}^{2/3} \nu_{15} H_{-11}
 \exp(-72.5 H_{-11} / T_3)   \nonumber
\hbox{~erg s$^{-1}$} ~~,
\end{eqnarray}
where the three terms on the right hand side correspond to X-ray
heating, radiative cooling, and sublimation cooling.  Here ${\cal Q}$ is the
total thermal energy in the grain, $T_3$ is temperature in units of
$10^3$~K, $\nu_{15}$ is the characteristic frequency $\nu_0$ in units
of $10^{15}$~Hz, and $H_{-11}$ is the binding energy per atom in units
of $10^{-11}$~erg.  The sublimation cooling (following Waxman \&
Draine 2000) is simply $H$ for each atom removed from the grain.

   As shown by equation~\ref{heatbal}, radiation and sublimation compete
to control the temperature.  Although both increase rapidly with increasing
temperature, sublimation has by far the more sensitive dependence on
temperature in this regime.  Consequently, radiation dominates at low
temperatures, sublimation at high.  Following Waxman \& Draine (2000),
we find the temperature that divides these two regimes by solving the
equation
\begin{equation}
T_{r=s} = { 2980 H_{-11} \over 1 - 0.041 \ln{ \left\{ {a_{0.1} \phi \over  
n_{23}^{2/3} \nu_{15} H_{-11} } \left[ T_{r=s} \over 2980 \hbox{K}\right]^5
\right\} } }~~\hbox{K}.
\label{requals}
\end{equation}
Note that we have introduced scaling factors into the logarithm so that
when the parameters all attain their fiducial values, $T_{r=s} = 2980$~K.
Because the typical $T_{r=s}$ is so high, for most of the volume of interest
radiation cooling dominates, as the burst is unable to heat the dust to
$T_{r=s}$.  The equilibrium temperature is then
\begin{equation}
T_{eq} = 1000\left[{A{\cal T}_h x_{K}^{1-\alpha} \over 
(2+\alpha)\phi t_{10}}\right]^{1/5} \hbox{K}.
\label{eq_temp}
\end{equation}
We have dropped the correction factor involving $x_{min}$ from this expression
because, except in the case of the very smallest grains, it does not 
substantially alter the result.  

    When $|da/dt| > a/t$ for grain size $a$ and X-ray emission
duration $t$, the grain is effectively destroyed by sublimation during the
burst and its aftermath.  If we approximate the X-ray lightcurve by
a square wave with this duration, the criterion for whether a grain
is completely sublimated is equivalent to a condition on its equilibrium
temperature:
\begin{equation}
T_{eq} \geq T_{\rm sub} = { 2360 H_{-11} \over
1 - 0.033 \ln(a_{0.1} n_{23}^{1/3} \nu_{15}^{-1} t_{10}^{-1} ) } ~~\hbox{K},
\label{subcrit}
\end{equation}
where $T_{\rm sub}$ is, of course, 
the temperature required to sublimate the dust
grain during the burst.
The critical temperature for sublimation is most sensitive
to the binding
energy per atom $H$, but also depends on grain size, density, and 
characteristic
sublimation rate $\nu_o$.  Comparing the result of equation~\ref{subcrit}
to equation~\ref{eq_temp}, we see that the flux ($\propto A/t_{10}$)
must be rather greater than our fiducial value in order to completely
sublimate most grains.

    Comparing the expressions for $T_{r=s}$ and $T_{sub}$, we also see that in
most cases the sublimation temperature is reached while cooling is still
radiation-dominated.  Sublimation cooling dominates radiative cooling
at temperatures below $T_{sub}$ only when
\begin{equation}
t_{10}  \le  1.7 \times 10^{-3} \phi^{-1} n_{23}^{-1/3} H_{-11}
\left(T_{\rm sub} \over 2980 \right)^{-5}~~.
\end{equation}
Sublimation cooling is therefore relevant only when the
burst is very short: $\la 0.02$~s.
This time is coincidentally comparable to the thermal
equilibration timescale for a grain at these temperatures ($\sim
10^{-3}$~s).  Thus, we do not examine closely the case
where the equilibrium grain temperature is set primarily by
sublimation cooling because any grain so close to the burst that this
regime applies will be destroyed well before the burst ends.

We have so far neglected grain temperature fluctuations, which can
matter because the sublimation rate is such a strong function of grain
temperature.  Such fluctuations may arise both from the rapid
substructure that characterizes most GRB light curves and from the
discrete nature of the X-ray heating.  All grains will be affected by
the light curve variations, while only grains with $a \la 0.01 \mu m$
will be substantially affected by the second mechanism.  To account
for such variations fully, one could calculate the temperature history
of a particular grain for a particular burst using
equation~\ref{heatbal} and so determine the total grain erosion $\int
da/dt(T) dt$.  Such a treatment is beyond the scope of this paper.
We simply note that for a typical burst, the fluctuations in heating
rate will result in more grain erosion than a uniform flux
$\propto A/ t_{10}$ would na\"{\i}vely produce.

Thus, for typical burst parameters, we find that the conditions under
which X-ray heating can vaporize both carbon-rich and silicate grains
are fairly similar.  In both cases, fluxes somewhat greater than those
associated with our fiducial parameter values are required.
Coatings of more volatile materials (ices, etc.)
can, of course, be removed much more readily.

    Waxman \& Draine (2000) pointed out that the optical/UV afterglow
can also heat the grains strongly enough to evaporate them.  Although the
peak flux in X-rays is often greater than in the optical/UV, grains absorb
a larger fraction of the optical/UV light striking them than of the
X-rays.  Consequently, which band is the more effective in sublimating grains
depends on details.  We contrast the two mechanisms in Figure~1 (see
discussion at the end of \S 3.3).

\subsection{Charging and electrostatic shattering}

    Although an electron that escapes does not contribute to grain heating,
it nonetheless can have a deleterious effect on the dust it leaves behind.
As the freed charge exits the grain, it leaves a positive charge,
and even if the X-ray flux in the burst proper is too weak to evaporate
grains along the line of sight, the build-up of electrostatic stress 
(Waxman \& Draine 2000) from these liberated electrons
can pose a lethal peril to the grains.  

     The rate of charging is simply the rate of photoionizations by
photons of high enough energy that the primary ionized electron has a
range greater than the size of the grain.  The majority of K-shell
photoionization events in medium-$Z$ elements lead to an immediate second
ionization by the Auger mechanism, but we make the conservative
approximation that the
energies of the Auger electrons are all below the threshold for escape.
Assuming that all individual grains are optically thin to X-ray photons, 
this rate is
\begin{equation}
R_{\rm ion} \simeq 2.2 \times 10^6 {A {\cal T}_i(x_{min},\alpha,N)
 \over 3 + \alpha}{a_{0.1}^3 \over t_{10}} x_K^{-\alpha}
\left(1 + x_{\rm min}/x_K\right)^{-(3+\alpha)} \hbox{~s$^{-1}$} .
\label{ion_rate}
\end{equation}
The function ${\cal T}_i$ is almost like ${{\cal T}_h}$.
The only difference is that the rate of charging is proportional to the
{\it number} of absorbed photons whereas the rate of heating is proportional
to the {\it energy} of absorbed photons, so the transparency factor ${\cal 
T}_i$
is weighted according to photon number rather than energy.  We defer
further discussion of both factors to \S 4.1.

   We emphasize that heating and charging depend on
the X-ray irradiation in different ways.  The grain temperature depends
on the X-ray {\it flux}, but the ultimate grain charge depends on the
X-ray {\it photon fluence}.  That is, because each sufficiently energetic
photon removes one electron, what matters is the total number of photons
in the energy range that, when absorbed in a grain, expel an electron.
This energy range is quite restricted.  The low-energy end is fixed by
the lowest K-edge plus $x_{min}$, and as the absorption cross section
drops rapidly with energy above threshold, the high-energy end is roughly
twice the highest energy K-edge plus
$x_{min}$. 
Consequently, grain
sublimation depends on the peak X-ray flux, whereas grain charging
depends on the integrated number of photons in the appropriate energy range.

   Two different mechanisms can control $x_{min}$: energy loss in the
grain and restraint by the same electrostatic potential that is built
up by charging.  Energy loss in the grain poses a threshold
$x_{min} \simeq  4 a_{0.1}^{2/3}$ for a grain with density
$\simeq 3$~gm~cm$^{-3}$.  This threshold for escape is great enough
that in most cases, charging is little affected by intervening absorption.
This is because there is a critical energy $x_c$ above which the 
ISM is effectively transparent, 
and typically $x_{min} + x_K > x_c$ (further
discussion of this point can be found in \S 4.1).
On the other hand, the electrostatic binding energy
of an electron at the grain surface is 
\begin{equation}
-eV = Z_g e^2/a = 1.4 \left({Z_g \over 10^5}\right) a_{0.1}^{-1} \hbox{~keV},
\end{equation}
where $Z_g$ is the grain charge in electron units and $e$ is the
electron charge (with a sign convention that $e = -|e|$).
The magnitude of the potential becomes larger than the local loss
threshold when
\begin{equation}
Z_g > 3 \times 10^5 a_{0.1}^{5/3} .
\end{equation}

    Unless the dust is in an environment with extremely high electron density, 
recombination with ambient electrons is far too slow to compete with burst
photoionization.  Allowing for the attractive focussing due to the grain
potential, the electron-grain cross section is
\begin{equation}
\sigma_{gr,e} = \sigma_{geom}\left(1 - {2 eV \over 3 kT_e}\right),
\end{equation}
where $\sigma_{geom}$ is the ordinary geometric cross section of the grain
and $T_e$ is the electron temperature.  When the grain potential is
large enough 
to create an interesting stress, $|eV| \gg kT_e$ unless the ambient 
electrons are
extremely hot.  Taking $\sigma_{geom} = \pi a^2$, we find that the
recombination rate is
\begin{equation}
R_{rec} \simeq 23 n_e a_{0.1} T_{e4}^{-1/2} \left({Z_g \over 10^5}\right)
\hbox{s$^{-1}$}
\end{equation}
for electron density $n_e$.   Comparing this rate to the rate estimated in
equation~\ref{ion_rate},
we see that electron recombination is unlikely to be important.  Although
the UV pulse associated with the burst itself might transform a dense
neutral region into a plasma with a high enough electron density for
recombination to be competitive, this can only occur much closer to the
burst than the critical distance for grain destruction by evaporative 
heating.

    Neither of the mechanisms controlling the energy threshold depends 
strongly
on the grain composition (only the density matters, and all the different
plausible compositions have similar density).  Consequently, $R_{ion}$
increases roughly in proportion to $(A {\cal T}_i) x_K^{3}$.  The positive
dependence on $x_K$ is due to the greater cross section when the photon energy
is closer to the threshold.  Although $\sigma_K$ is somewhat greater for
lower $Z$ elements, it does not change enough with $Z$ to outweigh
the proportionality of $R_{ion}$ to $x_K^{3}$.  Consequently,
higher-$Z$ compositions (e.g. silicates) are ionized more rapidly
than carbonaceous grains.  The contrast is about a factor of 10 in
the quantity $\sigma_{-19}n_{23}x_{K}^3$, from about 0.37 for pure C to
$\simeq 3.4$ for silicates.

    The electric stress $S$ created by this
potential depends on the shape of the grain, as well as on the mobility of
charges within the grain.  For example, in a perfectly conducting grain all
the charge would migrate to the outside.   If the grain were spherical, 
the physical stress would be symmetrically distributed around the grain.
However, sharp corners on a good
conductor will be loci of particularly large stress.  Insulating materials
would be likely to have more uniformly distributed stress, for the charge
density should simply be proportional to the initial K-shell electron density.
Unfortunately, our knowledge of the detailed shape and electrical properties
of grains is very shaky, and it is quite likely that grains exist across
a wide range of shapes and conductivity (see, e.g., Mathis 1998 for a
discussion of the issues involved).
Despite these uncertainties, we can at least make an order of magnitude
estimate of the stress:
\begin{equation}
S \equiv E^2/4\pi \sim (Z_g e)^2/4\pi a^4 = 1.8 \times 10^{10}
 \left({Z_g \over 10^5}\right)^2 a_{0.1}^{-4} \hbox{~dyne cm$^{-2}$} .
\label{chargestress}
\end{equation}

     To put this in context, it is necessary to estimate the tensile strength
of interstellar grains.  Much uncertainty also attaches to this number.  
Grains
with unflawed crystal structure could have tensile strengths $S_{crit}$
as high as
$10^{11}$~dyne~cm$^{-2}$ (Draine \& Salpeter 1979).  On the other hand,
impurities, lattice dislocations, and other imperfections could greatly
reduce the tensile strength.  Some (e.g., Burke \& Silk 1974) have suggested 
a value as low as $\sim 10^9$~dyne~cm$^{-2}$.  If grains are highly
porous structures (as suggested by Mathis 1996), the critical stress might
be still smaller.  For the current problem,
the large flux of energetic photons bombarding the grain is likely to
damage the grain's crystalline structure heavily during the course of
the burst.  Thus, the highest estimates of the
breaking stress are probably unrealistic in this context.  In the
following estimates, we will write the critical stress as
$S_{crit,10} \equiv S_{crit}/10^{10}$~dyne~cm$^{-2}$.

   For most grain compositions and for larger grains,
$x_{min} > x_K$ because, with the exception of Fe, all the elements
commonly found in grains have $x_K < 2$, whereas
$x_{min} \simeq 4 a_{0.1}^{2/3}$.
With that assumption, 
we find that when internal energy loss is the limiting factor,
the final charge of a grain is
\begin{equation}
Z_g \simeq \cases{1.1 \times 10^{5} A {\cal T}_i a_{0.1}^{1-2\alpha/3}
		\left(3 \over 3+\alpha \right) 4^{-\alpha} x_K^3
		& (General $\alpha$) \cr
	1.1 \times 10^5 A {\cal T}_i a_{0.1} x_K^3         & $\alpha = 0$ \cr
        1.9 \times 10^4 A {\cal T}_i a_{0.1}^{1/3} x_K^3   & $\alpha = 1$ 
\cr},
\end{equation}
resulting in a final stress
\begin{equation}
S \simeq \cases{
	2.1 \times 10^{10} (A {\cal T}_i)^2 a_{0.1}^{-2-4\alpha/3}
 		\left(3 \over 3+\alpha \right)^2 4^{-2\alpha} x_K^6
		   & (General $\alpha$) \cr
	2.1 \times 10^{10} (A {\cal T}_i)^2 a_{0.1}^{-2}    x_K^6
		   & $\alpha = 0$ \cr
        6.9 \times 10^{8}  (A {\cal T}_i)^2 a_{0.1}^{-10/3} x_K^6
		   & $\alpha = 1$ \cr}
\hbox{~dyne~cm$^{-2}$}.
\label{locstress}
\end{equation}

The above equations apply so long as local losses dominate
the energy requirement for electrons to escape the grain.  On the
other hand, if the grain charge becomes sufficiently large,
the energy required to escape the grain's electrostatic potential
exceeds the local losses.  Combining earlier results, we find that the
potential dominates if
\begin{equation}
A {\cal T}_i \ge 2.8 a_{0.1}^{2(1+\alpha)/3} \left(3+\alpha \over
3\right) 4^{\alpha} x_K^{-3} ~~.
\end{equation}
Equation~17 may be interpreted to mean that when the fluence accumulated
by a grain exceeds a minimum value, it enters the potential-limited
charging regime.  However, it is also possible that by that time it
has already accumulated a charge so great that it has been shattered.
For the criterion of equation~17 to be met, and yet for the grain
not to have been already destroyed, the stress given by equation~16
must be less than the critical stress.  Combining these two criteria,
we find that grains survive into the potential-limited regime only if
the critical breaking stress
\begin{equation}
S_{crit} > 2 \times 10^{11} a_{0.1}^{-2/3}\hbox{~dyne~cm$^{-2}$} .
\end{equation}
On this basis, we conclude that the potential-limited regime is
reached by only the biggest and strongest grains.

   When charging is limited by the electrostatic potential,
the eventual charge of the grain is
\begin{equation}
Z_g \simeq \cases{
	\left[ 8 \times 10^{21}  (1.4 \times 10^{-5})^{-\alpha}
	  \left(\alpha+4 \over \alpha+3 \right)
	  A {\cal T}_i a_{0.1}^{\alpha+6} x_K^3 \right]^{1/(4+\alpha)}
	  & (general $\alpha$) \cr
	3.2 \times 10^5 (A {\cal T}_i)^{1/4} a_{0.1}^{3/2} x_K^{3/4}
	  & $\alpha = 0$ \cr
	2.3 \times 10^5 (A {\cal T}_i)^{1/5} a_{0.1}^{7/5} x_K^{3/5}
          & $\alpha = 1$ \quad .\cr}
\end{equation}
Thus, when the grain charge is limited by the electrostatic potential
itself, it depends far more on grain size than on the strength of the
burst.   In this regime, the final stress is
\begin{equation}
S \simeq \cases{
	1.9 \times 10^{11} q_1(\alpha)
 (A {\cal T}_i)^{2 / (\alpha+4)} a_{0.1}^{-(4+2\alpha)/(4+\alpha)}
 x_K^{6/(\alpha+4)}		& (general $\alpha$) \cr
	1.9 \times 10^{11} (A {\cal T}_i)^{1/2} a_{0.1}^{-1} x_K^{3/2}
 & $\alpha = 0$ \cr
	1.0 \times 10^{11} (A {\cal T}_i)^{2/5} a_{0.1}^{-6/5} x_K^{6/5}
 & $\alpha = 1$ \cr}
\hbox{dyne cm$^{-2}$}.
\end{equation}
Here $q_1(\alpha) = \left[ 0.24^\alpha \left(3/
4\right)^{(\alpha+4)/4} {(\alpha + 4) / (\alpha + 3)} \right]^{2/(\alpha+4)}$
is a slowly varying function of $\alpha$ with value $q_1(0) = 1$.

We conclude that if $A > 1$, the stress should be great enough to break
any but the strongest and largest grains; most grains should crack
quite quickly.  Note that in all cases for which $x_{min} > x_K$, whether
the charging threshold
is set by local losses or by the grain potential, and for any reasonable
X-ray spectral slope, a shorter time is required to charge smaller grains to a
given stress.  A detailed discussion of the (relatively small) additional
fluence required to break grains down to very small sizes may be found in the
Appendix.

   Waxman \& Draine (2000) also discussed this mechanism of grain destruction,
but made several simplifying assumptions.  They supposed that there was a
uniform photoionization threshold of 10~keV and estimated the average
photoionization cross section per electron at $10^{-24}$~cm$^2$.  As a
result, our estimate of the grain destruction rate by this mechanism is
rather higher than theirs.

    We summarize all these results in Figure~1.
This figure shows the fluence $E_{51} D_{100}^{-2}$
required to either evaporate or shatter grains of size $a$ and zero
charge (i.e., in the notation of the Appendix, the fluence
corresponding to $t_{break}^0$). The spans in fluence and
grain size are chosen so as to display the widest plausible range
in these parameters while maintaining the validity of our physical
assumptions.  For example, grains smaller than
$\simeq 30$~\AA$ = 0.003 \,\mu m$ behave 
more like large molecules than grains; on the other hand, grains larger
than 1~$\mu m$ can be optically thick to some photoionizing X-rays, and
are no longer in the dipole limit when radiating infrared photons.  To
specify the curve positions, we set all scaling factors to unity, i.e.
${\cal T}_h = {\cal T}_i = 1$, $x_K = 1$, $S_{crit,10} = 1$, $t_{10} = 1$,
and $T_{sub} = 2300$~K.  The plotted curves account for the
inefficiency of X-ray heating in small grains due to the relatively
easy escape of fast electrons (eq.~\ref{range}).  They also use the
full forms of equations~2 and 9, i.e., they do not make the approximation
that $x_{min} \gg x_K$.  The comparison between
X-ray and optical/UV effects is fixed by using the observed fluxes of
GRB 990123 (Akerlof et al. 1999; Williams et al. 1999; E. Costa 2000,
private communication).  This burst was the fiducial burst chosen by
Waxman \& Draine (2000).  Because the X-ray spectrum of GRB 990123 is
not yet available, we have assumed a spectral index of $\alpha = 0$,
which is perhaps the most typical value for the bursts in the
BeppoSax sample (Frontera et al. 2000).   The UV evaporation rate
is calculated as in Waxman and Draine (2000),  but 
with the modification that the absorption
efficiency $Q_{UV} \equiv \min(1,2\pi a/\lambda_*)$ in order to
allow for the reduction in efficiency that occurs when grains
are smaller than the wavelength.  For the purpose of the figure,
we choose a characteristic wavelength $\lambda_* = 3000$~\AA.

   As the figure makes plain, many $\gamma$-ray bursts can eliminate large
parts of the dust-grain population.  For almost all reasonable parameters,
the electrostatic mechanism is a more powerful destroyer of grains
than is evaporation, whether due to X-rays or UV.  Silicate grains
are somewhat more sensitive to shattering than carbon-rich grains because
the cross section for X-rays energetic enough to expel electrons is
larger (the contrast is greatest for relatively large grains).  On
the other hand, evaporation acts more powerfully on carbonaceous grains
because they are able to absorb softer photons than silicate grains
can.

    The grain-size dependence of the critical fluence for
shattering is more complicated
to describe (the approximate scalings mentioned in this paragraph are
derived in the Appendix).  When $x_{min} \gg x_K$ is a good approximation,
the critical fluence declines with decreasing $a$, roughly
$\propto a^{1 + 2\alpha/3}$.  This approximation is best for large
grains; the dividing line between ``large" and ``small" for this
purpose falls at smaller $a$ for carbon-rich grains than for silicates.
On the other hand, when $x_{min}$ is not much larger than $x_K$, the
critical fluence grows with decreasing $a$, roughly $\propto a^{-1}$.  Thus,
particularly when exposed to soft X-ray spectra, small carbon-rich
grains are more readily destroyed than large ones.  The 
same trend applies to silicate grains, but more of parameter space
is occupied by ``small" grains for which $x_{min}$ is not large
compared to $x_K$; ``small" silicate grains may require as much or
more fluence to break as ``large" ones, particularly when the X-ray
spectrum is very hard.  The critical flux for X-ray evaporation is relatively
insensitive to grain size because both heating and cooling rates
scale approximately $\propto a^3$.

   For all three values of $\alpha$, the critical fluence dividing the
potential-limited grain-charging regime from the local loss-limited regime
is significantly greater than the critical fluence for shattering grains.
This fact means that, at least
when our fiducial parameters are appropriate
(most importantly, for $S_{crit,10} = 1$),
grains shatter before their charge becomes so great that the potential
becomes important.

If electrostatic stresses are, for some reason, ineffective, evaporation
can also destroy grains, but for a smaller range of parameters.  If the
optical/UV to X-ray flux ratio of GRB 990123 is representative of most bursts,
UV heating is generally the more important effect for small grains, X-ray
heating for large (although the dividing line depends on grain composition).
The primary reason for this is that the absorptive
efficiency of dust grains is fairly high in the optical/UV band, but
falls rapidly with increasing energy through the X-ray band.  All but
the thickest dust grains are optically thin throughout almost the
entire relevant range of X-ray energies.  


\begin{figure}
\centerline{\psfig{file=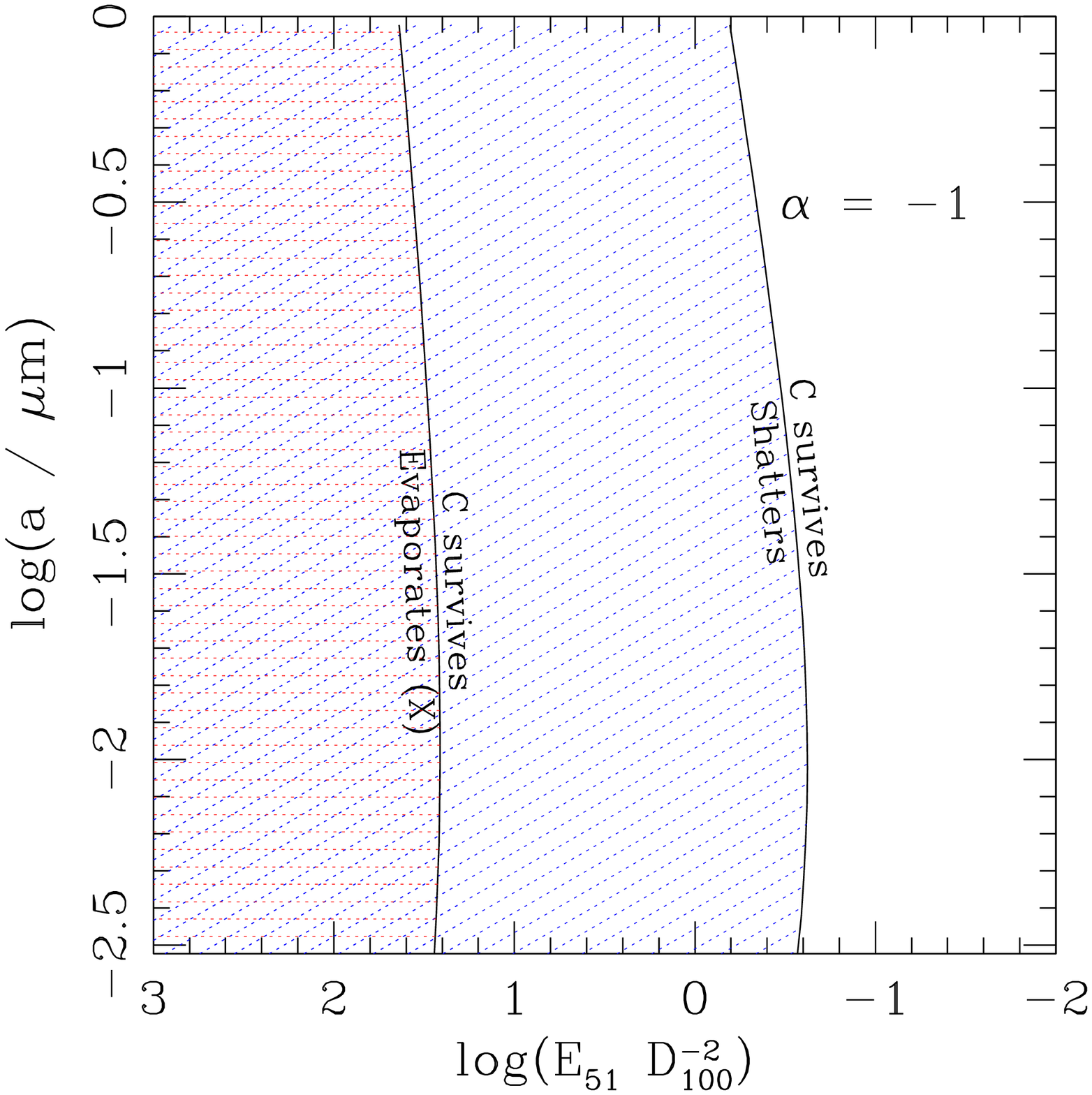,height=2.45in,width=3.0in}
            \psfig{file=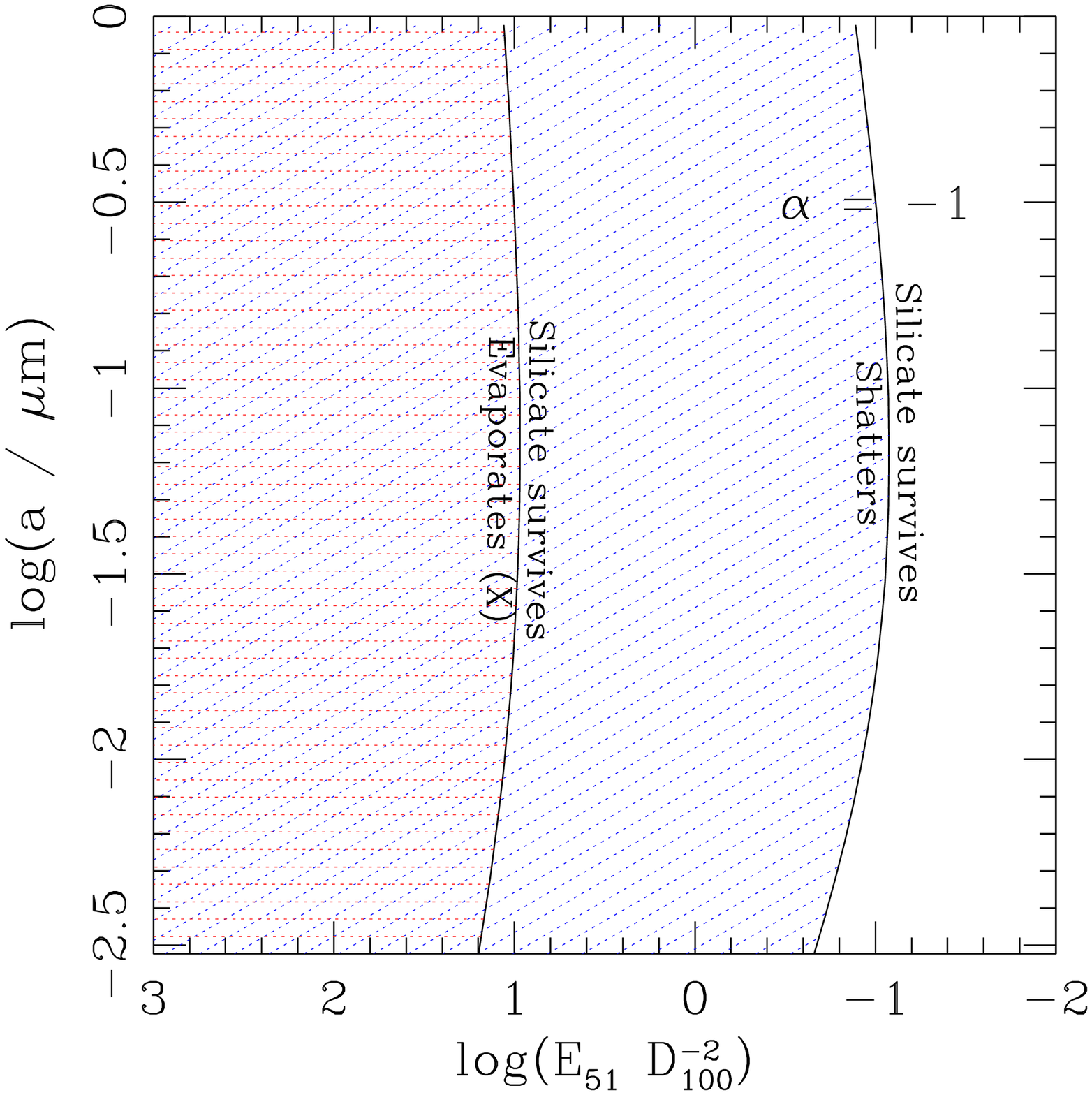,height=2.45in,width=3.0in}}
\centerline{\psfig{file=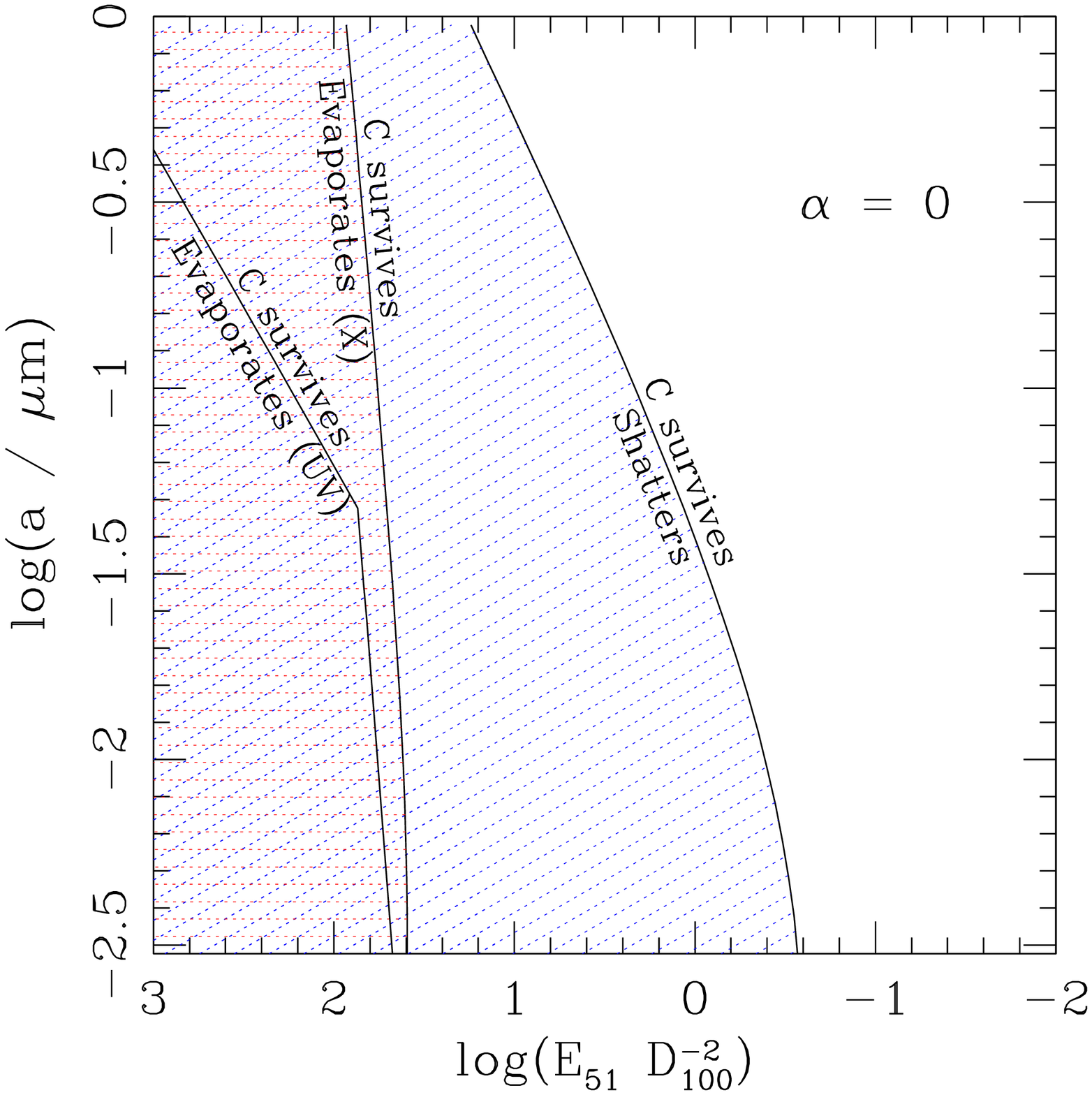,height=2.45in,width=3.0in}
            \psfig{file=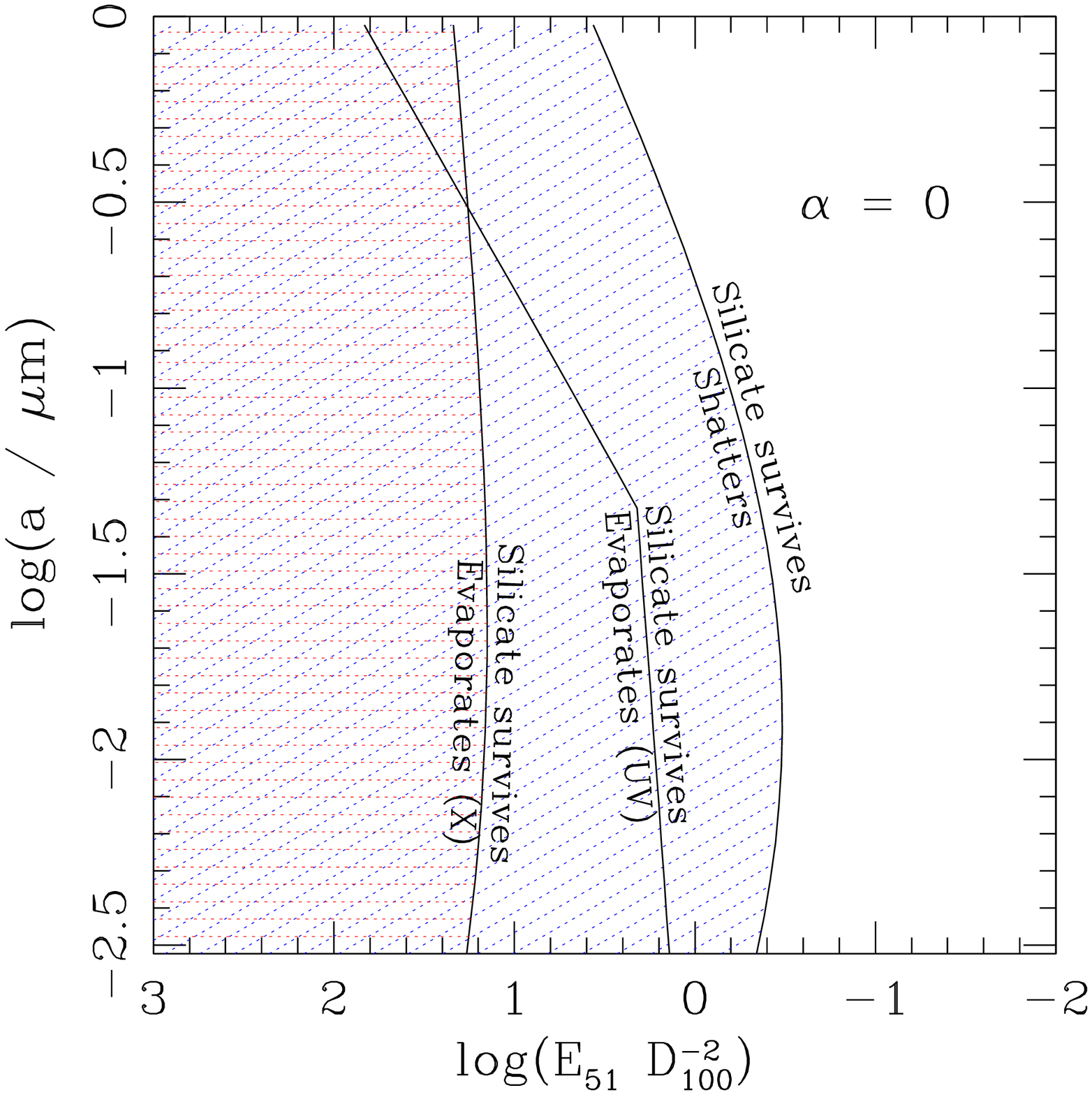,height=2.45in,width=3.0in}}
\centerline{\psfig{file=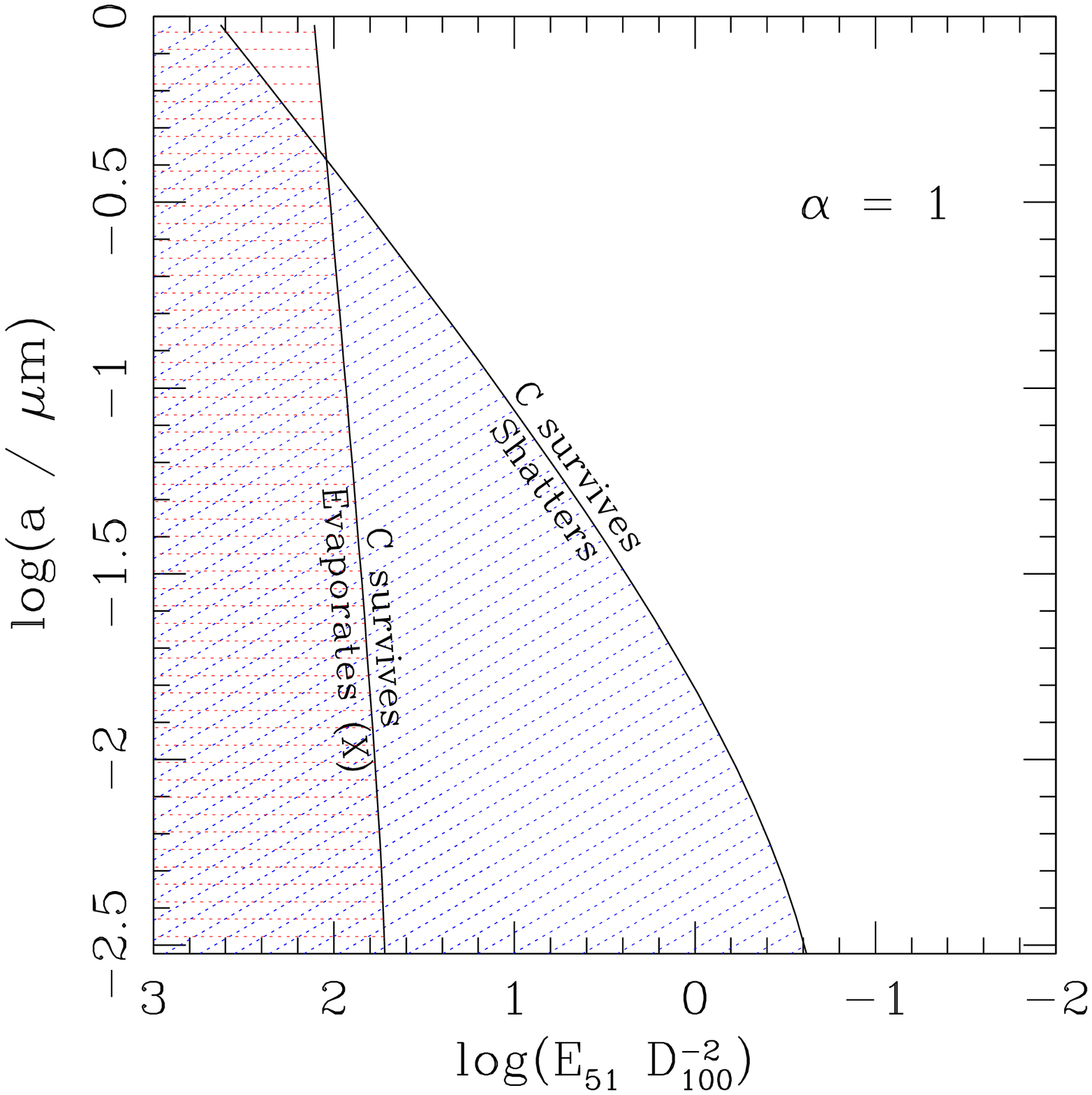,height=2.45in,width=3.0in}
            \psfig{file=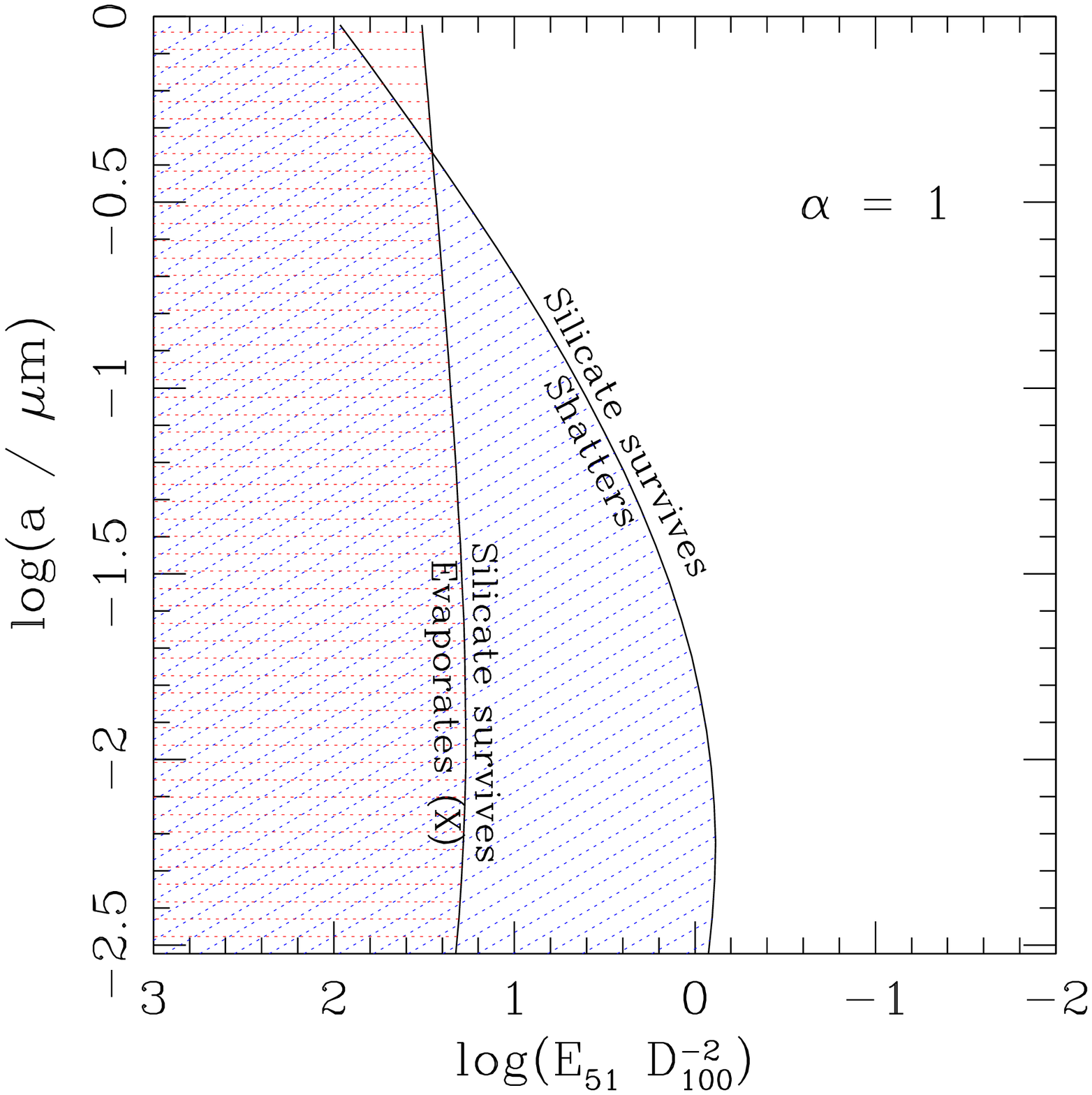,height=2.45in,width=3.0in}}
\caption{
Figure 1: Grain destruction thresholds as a function of scaled
GRB fluence $E_{51} D_{100}^{-2} {\cal T}$ and grain radius $a$.  The
six panels correspond to the X-ray spectral indices $\alpha = -1$, $0$,
and $1$ for carbonaceous grains (left side) and silicate grains (right side).
Other relevant parameters were taken at their fiducial values
($S_{10} = 1$, $t_{10} = 1$, $\phi$ as given by Draine \& Lee 1984 for
the respective grain compositions); changes to these would affect
the relative importance of the different destruction mechanisms
considered (see equations).  All panels show thresholds for (1) grain
shattering by electrostatic stress; and (2) grain evaporation by X-ray
heating.  In the $\alpha=0$ case, we also show the threshold for (3)
grain evaporation by UV and optical heating. }
\end{figure}

\section{Grain destruction distances}

   In the simplest circumstances the maximum distance from the burster at 
which
each mechanism will be effective may be read off the curves in the figure.
In this section we provide further guidance on estimation of these
maximum distances: we discuss how those curves may change due to intervening
absorption, and we present some analytic approximations to estimate
the critical destruction distances.

\subsection{Intervening absorption}

   Medium-$Z$ atoms in the ISM of the host galaxy
between the burst source and the dust may intercept X-rays before they
strike the grains.  Whether these atoms are able to do so depends 
on whether or not they are fully stripped of electrons; the specific
ionization stage of the atoms affects the magnitude of their photo-electric
opacity to only a modest degree.

   In the energy range between 538~eV (its edge when neutral)
and 7.1~keV (the neutral Fe K-edge), O provides the single greatest
contribution to X-ray photoionization opacity, so we concentrate on
estimating the circumstances in which it may be fully ionized.  As its
ionization stage rises from neutral to H-like, the K-edge of O
rises from 538~eV to 871~eV, with most of the change occurring between
the Li-like and H-like stages.  The cross section just above threshold
varies relatively little, dropping from $\simeq 5 \times 10^{-19}$~cm$^2$
for OI to $\simeq 1.4 \times 10^{-19}$~cm$^2$ for OVIII (Verner \& Yakovlev
1995).  For each K-shell ionization in OI through OV, there is (on
average) almost one Auger ionization.  Therefore,
in order for O to be fully-stripped by the burst, a number fluence of
photons above 500--800~eV $\gtrsim 2 \times 10^{19}$~cm$^{-2}$ must pass
through the region where the O is located.  Scaled in terms of our fiducial
parameters, the photon fluence above energy $x_K$ in a burst is
$5 \times 10^{17} x_{K}^{-\alpha} E_{51} D_{100}^{-2}$~cm$^{-2}$.  Thus,
we expect that closer than $\sim 6$~pc, a typical burst releases
enough X-ray photons to strip all the O atoms; farther away, the O
may be partially, but not completely, ionized.  Recombination takes
place so much more slowly (on a timescale $\sim 10^{11} n_{e}^{-1}$~s)
that it is irrelevant to burst afterglows.

   If the ISM has Solar abundances and the bulk of the absorbing column
is located far enough from the burst that O (and similar atoms) are not
fully ionized, the soft X-ray opacity for photons of energy
$0.5 \lesssim x \lesssim 5$
scales roughly $\propto x^{-3 + \delta}$ where $\delta \simeq 0.4$ -- 0.6
depends on the ionization state (Morrison \& McCammon 1983, Krolik \&
Kallman 1984).  The departure from $x^{-3}$ scaling is due to the
summation of photoelectric opacity from many different elements, each
having a different threshold energy (in contrast to the situation for dust,
in which a few elements dominate the opacity of any particular grain).
As a result there is effectively an 
energy $x_c$ below which the ISM is optically thick, but above which it
is transparent.  The transparency
factors ${\cal T}_{i}$ and  ${\cal T}_h$
can then be evaluated by appropriate integrations:
\begin{equation}
{\cal T}_h \simeq \cases{1 & $x_c < x_K$ \cr
                  \left(x_K/x_c\right)^{2+\alpha -\delta} & $x_K < x_c < x_K + 
x_{min}$ 
\cr
                  0                               & $x_c > x_K + x_{min}$\cr}
\end{equation}
and
\begin{equation}
{\cal T}_i = \cases{1 & $x_c < x_K + x_{min}$ \cr
                    \left[\left(x_{min} + x_K\right)/x_c\right]^{3+\alpha 
-\delta}
                      & $x_c > x_K + x_{min}$ \cr}.
\end{equation}
In effect, when $x_c > x_K$, $x_c$ replaces $x_K$ in the heating rate; 
similarly,
when $x_c > x_K + x_{min}$, $x_c$ replaces $x_K + x_{min}$ in the ionization
rate.

    In the ISM of an ordinary galaxy, typical total column
densities are $N \sim 10^{21}$~cm$^{-2}$; however, compact star-forming 
regions
may have column densities as much as 100~times greater.  If the gas
is nearly neutral,
random lines of sight might have $x_c \simeq 0.55$ while lines of sight
passing through giant molecular clouds or star-forming regions might have
$x_c$ as large as $\simeq 3.4$.

    Because lower-energy K-shell ionizing photons contribute primarily to
heating, whereas higher-energy photons (those above $x_K + x_{min}$)
contribute to ionization, intervening absorption can more easily affect
heating than charging.  Its effect can be particularly great on
carbon-rich grains, whose $x_K$ is much lower than in silicate grains.

    However, as the figure demonstrates, we find that grain charging is the
dominant grain destruction mechanism under almost all circumstances.   Because
$x_{min}$ is usually at least a few keV (except for very small grains),
soft X-ray absorption is most likely to affect the charging rate only for
small, C-rich grains unless the absorbing column is relatively great.
Even in those instances in which the column is larger than average,
it is possible that most of the intervening matter is associated with
a molecular cloud close enough to the burst that its C, N, O, etc.
are fully-ionized by the burst itself.

   In the discussion so far we have not made any distinction between
medium-$Z$ atoms in grains or in the gas phase.  This is because their
K-shell opacity does not depend at all on the phase in which they
are located.  Consequently, unlike the situation regarding UV propagation,
dust destruction is irrelevant to the soft X-ray opacity.  

\subsection{X-ray heating}

We now estimate the destruction distance due to X-ray heating.
The maximum distance for dust evaporation by this mechanism 
follows easily from
equations~\ref{eq_temp} and~\ref{subcrit}.  The condition for
grain evaporation is
\begin{equation}
D_{100} \le 0.11 \left\{ { C {{\cal T}_h} x_k^{1-\alpha} \over
 (2+\alpha) \phi t_{10} H_{-11}^5 } \left( 1 - 0.0326 \ln\left
 [ a_{0.1} n_{23}^{1/3} \over \nu_{15} \; t_{10} \right] \right)^5
 \right\}^{1/2}
\end{equation}
where $C = E_{51} \sigma_{-19} n_{23} = D_{100}^2 A$.
So long as the logarithm on the right hand side is small, this
expression is well approximated by
\begin{equation}
D_{100} \la 0.11  \left\{ { C {{\cal T}_h} x_k^{1-\alpha} \over
 (2+\alpha) \phi t_{10} H_{-11}^5 }
 \right\}^{1/2} \left\{ a_{0.1} n_{23}^{1/3} \over \nu_{15} t_{10}
 \right\}^{-0.0815}~~.
\end{equation}

    Two effects not contained in these equations will alter the grain
destruction distance for small grains in opposite ways. 
First, as already remarked, small grains are heated less efficiently
than large ones because it is easier for electrons to carry off
the deposited energy.  Second, we have used the
equilibrium temperature.  Thermal fluctuations due to the
discreteness of the impinging X-ray photons can be substantial for very
small grains, and may allow sublimation of these grains even when
their equilibrium temperature is appreciably below $T_{sub}$.
The characteristic values of $E_{51}$ should range from $\sim 1$ to
$\sim 10$ based on BeppoSAX observations of GRB 970228, 970508, and
971214 (Frontera et al 1999), while $t_{10}$ ranged from $\sim 3$ to
$\sim 5$ for these same bursts.  Thus, destruction distances in the range
of 5 to 20 parsecs are reasonable for this mechanism.

\subsection{Grain charging}

For grain shattering through charging, the destruction distance
can be calculated separately for the cases where the energy
requirement for an electron to escape the grain is dominated by
internal energy losses or by electrostatic potential.  The appropriate
destruction distance will be the smaller of the two.  Here we present
several analytic approximations for the destruction distance, all
valid for the regime in which $x_{min} \gg x_K$.  For the case
of internal energy losses, we find
\begin{equation}
D_{100} \le 1.2 \left(3 \over 3 + \alpha \right)^{1/2}  4^{-\alpha/2}
 \sqrt{ C {\cal T}_i } \, S_{crit,10}^{-1/4}
 a_{0.1}^{-1/2 - \alpha/3} x_K^{3/2} .
\end{equation}
For the case where electrostatic potential determines the minimum
photon energy for charging, we instead find
\begin{equation}
D_{100} \le 19 q_2(\alpha) \sqrt{ C {\cal T}_i } \, S_{crit,10}^{-1 - 
\alpha/4}
 a_{0.1}^{-1-\alpha/2} x_K^{3/2} ~~,
\end{equation}
where $q_2(\alpha) = 1.02^\alpha \left[ (3/4)^{1+\alpha/4} 
(\alpha +4)/(\alpha +3) \right]^{1/2}$.
Thus, this mechanism can yield destruction distances considerably
beyond those offered by evaporation; as the figures show, the contrast
can be anywhere from a factor of 3 to a factor of 10.  Note, however, that 
there
is a hidden dependence on grain size and composition through the
dependence of ${\cal T}_i$ on $x_{min}$ and $x_K$ (see Eqns. 21 and
22).  
${\cal T}_i$ decreases for smaller $x_{min}$ and $x_K$, 
and $x_{min}$ decreases for smaller $a$, so that soft X-ray
absorption becomes more important for smaller and more C-rich grains,
diminishing the maximum distance at which they can be destroyed.

\subsection{UV heating}

Finally, we compare these results to the destruction radius for UV
evaporation of grains.  Waxman \& Draine (2000, eqn.~17) obtain a destruction
radius of
\begin{equation} 
D_{100} \approx 0.12 \sqrt{ L_{49} (1 + 0.1 a_{0.1}) \over a_{0.1} }
\end{equation}
where $10^{49} L_{49}$~erg~s$^{-1}$ is the total burst luminosity
in photons with energies
between $1$ and $7.5$ eV.  Waxman \& Draine estimate $L_{49} \approx
1$ based largely on the observed prompt emission from GRB 990123,
whose optical flash now appears to have been brighter than average
(Williams et al. 1999, Akerlof et al. 2000).
Thus, heating by the UV-optical flash and by X-rays have roughly comparable
importance for grain evaporation, and either may be more important,
depending on the burst spectrum and grain size.  Neither evaporation
mechanism competes with grain shattering under normal circumstances.

   At this point one may raise the question, ``How is it that charging
is able to destroy grains so much farther away from the burst than
heating can?  In the end, both mechanisms must supply the energy to
break the same number of bonds."  The answer to this question has
several components.  First, shattering to small sizes may still leave
the grains as clusters of $\sim 100$ atoms; if so, in contrast to
sublimation, which transforms the grain into individual free atoms,
only $\sim 1\%$ of the bonds have been broken by the charging
mechanism.  Additionally, shattering exploits existing defects in grain
crystalline structure, again reducing the number of bonds that must be broken
to split the grain.  Finally, both mechanisms have small,
but different, efficiencies.  Energy is wasted in the charging
mechanism by giving free electrons more energy than is required to
leave the grains and by the energy lost to heat when ionizations
occur that don't result in electrons escaping the grain.  In the
case of radiative heating, even more energy is lost by radiative
cooling.  Thus, the contrast in effectiveness of these two
mechanisms is due to the quantitative balancing of several different
effects.

\section{Discussion: Burst After-Glow Extinction}

   Grains evaporated by the heating mechanism immediately cease to
contribute anything to extinction because their material is dispersed
in the gas phase.  Laboratory simulations of sudden grain heating have
shown that they evaporate into atoms and small molecules such as SiO
(Duley \& Boehlau 1986).  On the other hand, when a grain is shattered by
electrostatic stresses, the immediate effect is to create several new,
smaller grains.  As long as these pieces are large enough to act effectively
as ``solids", their total absorptive opacity is unchanged, but their
total scattering opacity for long wavelengths is sharply diminished
while their total scattering opacity for short wavelengths is increased.
However, this immediate result of shattering is not the final
result.  As discussed in the Appendix, the total time required for
successive grain shatterings to pulverize them down to very small sizes
is only a modest multiple of the time required for the first break.

    All of the estimates so far depend on treating the grains as
macroscopic solids whose intrinsic properties are independent of size.
However, as they are broken into smaller pieces,
this assumption becomes questionable.  It is possible, for example,
that dislocations or fractures become increasingly rare as larger
grains are split into smaller pieces, in part by cracking at defect sites.
If so, the critical stress would increase as $a$ decreases.  On
the other hand, it is also possible that the continuing irradiation
by the burst and its afterglow---and the shattering events themselves---may
create new lattice defects.
    
\subsection{Atomic cluster opacity}

   One possible outcome, then, is that rising critical stress
causes the fragmentation process to slow, and perhaps stop, as the grains
reach a typical size of a few tens to hundreds of \AA.  The physics of such
``atomic clusters" is unfortunately very poorly understood because
neither the approximations appropriate to macroscopic solids nor
those appropriate to molecules are valid in this transition regime.

    We can, however, make some rough guesses about the optical
behavior of a population of such small grains.  In this case, 
the effect of dust destruction on extinction 
should vary strongly with wavelength because of the
differing dependence of scattering and absorption efficiencies on grain
size.  Because the absorption efficiency of grains smaller than a wavelength
scales $\propto a/\lambda$, the total absorption opacity depends
only on the mass in grains, and
not at all on the grain size distribution; scattering opacity, however,
depends strongly on grain size.  At rest-frame wavelengths greater than
$\simeq 2200$~\AA, the dust albedo (assuming a conventional dust size
distribution) is $\simeq 0.8$ (Laor \& Draine 1993); consequently,
for these wavelengths, grain shattering reduces the total extinction by
about a factor of 5.  Between $\simeq 2200$~\AA\ and $\simeq 1400$~\AA\
the albedo falls to $\simeq 0.4$, and is roughly constant from there
to $\simeq 80$~\AA.  Therefore, shattering grains to the atomic cluster scale
reduces the extinction for $\lambda < 1400$~\AA\ by only
about a factor of 0.6.

\subsection{Large molecule opacity}

    Should the shattering process continue to operate on atomic
clusters, the ``grains" begin to behave more like large molecules
than small solids.  The issue of critical stress gradually changes to
the issue of the existence of bound states for highly-charged
molecular ions.  UV photons, although relatively
ineffective at causing electron losses in larger grains because
they cannot penetrate the surface, become more efficient at ionization
when the structure is only a few tens of \AA\ across.
The absorptive opacity, which varies
comparatively smoothly with wavelength in solids, gradually becomes
concentrated into a number of molecular resonances.  In this
subsection we discuss the effect on inter-stellar opacity if
grain-shattering results in a new population of large molecules.

      It is now widely thought that a substantial fraction of the UV
extinction in our own galaxy may come from polycyclic aromatic hydrocarbons
(PAHs) (cf. Li and Draine 2001).  These large molecules (in some
cases made up of more than a dozen benzene rings) may be a by-product
of the destruction of carbonaceous grains.  As absorption by PAHs is
now suspected to be the primary cause of the $2200$~\AA\ feature
in the Galactic extinction curve (Weingartner and Drain 2001),
this feature may remain even as the carbonaceous grains are destroyed.
However, there is substantial evidence that  the ionizing
radiation of both AGN (Rigopoulou et al. 1999, Laurent et al. 2000)
and young stars (Boselli et al. 1997) can destroy PAHs.  Indeed,
studies of extinction in starburst galaxies find  no evidence for
PAH-induced absorption (Gordon, Calzetti and Witt, 1997, Calzetti, Kinney
and Storchi-Bergmann 1994).  It is therefore doubtful that the PAHs
formed by the shattering of large grains could long survive the UV
radiation from the burst without protection from intervening larger grains.
Thus, while a PAH feature might appear during the burst, and mid-IR emission
from the excited PAHs might be visible (Bauschlicher and Bakes 2000),
these particles might not survive to provide long-lasting attenuation.

\subsection{Small molecule opacity}

   When silicate grains are destroyed, there is no large molecule analog
of PAHs; instead, they may be broken into small molecules such as SiO.
These molecules offer a rich UV line spectrum, and substantial enhancement
of their abundance as a result of returning heavily-depleted 
Si to the gas phase may sharply increase the equivalent width
of these lines.  For example, in the $E \, ^{1}\Sigma^+$ -- $X \, 
^{1}\Sigma^+$
system of SiO there are band-heads at 1924.6~\AA, 1900.2~\AA,
1876.7~\AA, $\ldots$ (Elander \& Smith 1973).  There is another potentially
observable system ($J \, ^{1}\Pi$ -- $X ^{1}\Sigma^+$) whose 
longest-wavelength
band-head is at 1310.1~\AA\ (Morton 1975).  In other molecules (e.g.,
CO: van Dishoeck \& Black 1988), some electronic transitions lead to
photodissociation; if transitions like these also exist in SiO, UV
absorption bands might appear for only as long as it takes the
continuing UV afterglow to destroy the newly-created molecules.  
Unfortunately,
even for known bands the oscillator strengths are very uncertain
(Pwa \& Pottasch 1986), so it is difficult to make quantitative predictions
for these features.

    It is conceivable that if carbonaceous grains are broken into PAHs
and silicate grains are transformed into SiO and related molecules,
there might then be so many strong
molecular lines as to recreate an effectively continuous opacity; however,
the most astrophysically abundant elements found in grains (i.e. C and O)
are not depleted by more than a factor of a few anywhere, while the next
most abundant (Mg and Si) are never depleted by more than a factor of 10
(Savage \& Sembach 1996).  Consequently, the additional opacity
created by molecules formed from destroyed grains is unlikely to increase
the effective continuous opacity by a large factor.

\subsection{Atomic and ionic opacity}

    If grain-shattering releases individual atoms and ions into the ISM,
species that are ordinarily strongly depleted may produce observable
features.   For example, in cold regions, the depletion factors for
Ca, Ti, Cr, Fe, Co, and Ni range from $\sim 10^2$ to $\sim 10^4$; even
in warmer regions they are depleted by factors of $\sim 10$ -- 100
(Savage \& Sembach 1996).
If grains are truly broken all the way to atoms and ions, these elements
are returned to the gas phase, where their ability to produce absorption
lines is enhanced relative to the normal circumstance in which
they are locked into solid particles.  Because these elements also have
first ionization potentials smaller than that of O, the most abundant
element in silicates (6 -- 8~eV rather than 13.6~eV), one might expect
that they will take up a disproportionate share of the charge.  Therefore,
immediately after the destruction of the grains these elements would be
a mix of neutral atoms and singly-charged ions.

     The ideal lines to use as signatures of dust break-up are resonance
lines in these species that fall in the rest-frame UV (so that
redshifts of 1 -- 2 make them easily observable) and have undepleted
optical depths $\sim 1$--100 (so that they are easily detected if undepleted,
but weak normally).  The optical depth at line center is
\begin{equation}
\tau_{lc} \simeq 1500 f N_{21}
                      \left({\lambda \over 1000\hbox{~\AA}}\right)
                      \left({b \over 10\hbox{~km~s$^{-1}$}}\right)^{-1}
                      \left({X_i \over 10^{-5}}\right),
\end{equation}
where $f$ is the oscillator strength, $N_{21}$ is the H column density
in units of $10^{21}$~cm$^{-2}$, $\lambda$ is the wavelength of the
specific transition, $b$ is the characteristic velocity width of the
atoms, and $X_i$ is the population relative to H of the ground state
of the transition.  From this expression we see that the lines capable of
signalling dust break-up may vary from case to case (depending,
for example, on $b$), but in many cases will have oscillator
strengths rather less than unity.

     Here we present a few examples of candidate lines.  Because of the large
number of variables that might affect any particular burst, we
stress that there is no way to select a list of lines that will,
in all circumstances, be the most likely to appear only when the grains
along the line of sight are shattered.  We
concentrate on the ions \ion{Ca}{1}, \ion{Ca}{2}, \ion{Ti}{1},
and \ion{Ti}{2} because their ground states are sufficiently well-separated
in energy from other levels that one might reasonably expect them to
contain most of the population of these ions; the spectroscopy of neutral
or singly-ionized Fe-group elements is so much more
complicated that there is no simple way to estimate the fractional population
in the lower level of any particular transition.  Governed by these
considerations, we have chosen, from among the many listed in Morton (1991),
a few lines that might be particularly promising:
\ion{Ca}{1}~2275;
the 8-line multiplet \ion{Ti}{1}~2934--2951;
the 11-line multiplet \ion{Ti}{2}~3217--3238;
and the 10-line multiplet \ion{Ti}{2}~1895--1911.
If all of each element's population were in the
ground state of the species in question,
the local abundances relative to H were as given by Anders \& Grevesse (1989)
for the Solar System,
and $N$ and $b$ had their fiducial values,
the undepleted line-center optical
depths for these transitions (for each line in a multiplet, but averaging
over the lines within a multiplet)
would be 48, 3.8, 8.5, and 2.4 respectively.
Even the depletion seen in warm regions of our own Galaxy's 
ISM would cause these lines to disappear, but return to the gas
phase might return them to observability.  If a burst were observed very
soon after it begins, it might even be possible to watch a line of this
sort grow in strength.

\section{Conclusions}

     This  work demonstrates that, at least under certain circumstances,
X-rays emitted during and immediately after a $\gamma$-ray burst
can substantially diminish extinction along the line of sight to a burst.
However, as shown by both the numerous scaling parameters we employ
and our discussion of the physics uncertainties, this does 
not necessarily happen.  Given the wide dispersion in the properties of burst 
afterglows, not to mention the likely wide dispersion in the distribution of
dust along the lines of sight to bursts, all possibilities, from null to 
complete elimination of extinction may happen. 

    Because the X-rays are largely emitted within tens of seconds to 
minutes after the start of a burst,
yet the optical transient may be observable well before the X-rays cease, in
some cases it may be possible---if a response time $\la 1$ minute can
be achieved--- to watch the extinction change as the grains are
destroyed.  Indeed, both the strength and color of the extinction
may vary, as larger grains are pulverized down to smaller grains,
and as molecules and atoms are released from the dust, and sometimes
ionized.  We have identified several candidate spectral features that
might signal these processes.

    We close by noting that the physics discussed above has potential
applications to other astrophysical objects.  Soft gamma repeaters
may destroy dust locally by the same mechanisms as GRBs.  Quasars are
likely to destroy dust through charging and electrostatic shattering,
although their lower peak fluxes render X-ray heating and sublimation
fairly ineffective except very close to the galactic nucleus.
On a smaller scale,  Galactic ``microquasars''
and X-ray binaries may similarly shatter nearby grains.  These
objects, with their persistent or repeating emission, may offer
another useful laboratory to search for X-ray grain destruction.

\acknowledgments

    We thank David Neufeld for instruction in molecular spectroscopy,
Daniela Calzetti and Sangeeta Malhotra for discussions of the
properties of interstellar dust, and our second referee, John Mathis,
whose insightful report led to significant improvement in the clarity of this
paper.  

    This work was partially supported by NSF Grant AST-9616922 and NASA Grant
NAG5-9187 (JHK). JER's work is supported by an
Institute Fellowship at The Space Telescope Science Institute
(STScI), which is operated by AURA under NASA contract NAS 5-26555.

\vskip 0.5cm

\appendix
\section{Appendix: The Time Required to Break Grains to Molecular Size}

     As shown in Figure~1, the fluence required to break grains generally
decreases with diminishing size over much of the size range between
1 and 0.01~$\mu$m, and then slowly increases as the grain size
becomes still smaller.  Consequently, the total fluence necessary
to break ``large'' grains into pieces $\sim 0.01$~$\mu$m in size is
a modest multiple of the fluence accumulated before the first break
provided the critical stress does not increase greatly as $a$ decreases.
The total fluence absorbed in order to crack grains into still smaller
pieces is also comparable to the fluence that produces the first break
(given the shakier assumption that ``solid state'' behavior applies)
because the dynamic range in size between the
scale on which the curves ``bend'' and the molecular scale is not too great.
The remainder of this Appendix will be devoted to quantifying these
statements.

     We begin by estimating the total fluence required to traverse
the first part of the curve.  The fundamental physical reason for the
change in trend is the transition from $x_{min} > x_K$ to $x_{min} < x_K$.
Consequently, to make this first estimate we assume that
$x_{min} > x_K$.  To further simplify the calculation, we also
make the approximations that grains
break into $k$ round pieces each time they are electrostatically shattered,
and that they share the charge evenly.  It is conceivable that in grains with
good electrical conductivity (e.g., those made of graphite) the charge
would all migrate to the surface, and that this surface layer might break
off, leaving a remnant almost as large as the original grain.  However,
in addition to the requirement on the electrical conductivity, this
process would also require any imperfections in the grain lattice structure
to concentrate near the surface.  We view this combination of prerequisites
as unlikely.  Finally, for convenience of discussion, we also assume that
the X-ray flux is constant during the time that it is ``on'', so that time
is proportional to fluence.

    From Equation~\ref{chargestress}, we see that there is a critical charge
for shattering grains,
\begin{equation}
Z_{g,crit} = 7.5 \times 10^4 a_{0.1}^2 S_{crit,10}^{1/2} .
\label{critcharge}
\end{equation}
Combining all the scaling factors except those involving $x_K$ and $a$ into
one constant $D$, the time to reach this critical charge starting from $Z_g = 
0$
is then
\begin{equation}
t_{break}^0 \simeq 4^{3+\alpha} D x_{K}^{-3} a_{0.1}^{1 + 2\alpha/3}.
\end{equation}
So long as $\alpha > -3/2$ ($-1 < \alpha < 1$ is the usual range),
$t_{break}$ does indeed decline with decreasing $a$.  We now rewrite the
breaking time in terms of how many shattering events were required to reach
that size.  Using the assumption of even breaking into round pieces,
the size of the fragments after $m$ breaks is $a_m = a_o k^{-m/3}$
(with $a_o$ the original grain size).  If
these fragments were born with no charge, the X-ray irradiation time
to break them the $m+1$-th time would be
\begin{equation}
t_{break}^m \simeq t_{break}^0
            k^{-m(1 + 2\alpha/3)/3} .
\end{equation}

    In fact, the true time to reach the breaking point is somewhat smaller
because each piece inherits its share of the charge on the original grain.
Using the relation $a_m = a_o k^{-m/3}$ in Equation~\ref{critcharge}, we see
that the charge per fragment required to produce the $m+1$-th break
is $\propto k^{-2m/3}$.  If
this is divided evenly among the successor fragments, each one acquires
a charge $\propto k^{-2m/3 - 1}$.  But the charge required to shatter this
next generation is $\propto k^{-2(m+1)/3}$, so the fraction of the required
charge with which the fragments are born is $k^{-1/3}$.  Thus, the fluence
necessary to reach the critical stress the next time is reduced by the
factor $1 - k^{-1/3}$, a quantity that could be anywhere from, say, $\simeq 
0.2$
(for $k=2$) to $\simeq 0.5$ (for $k=10$).

     Combining this correction factor with the estimate for $t_{break}$, we
find that the total time between the initiation of the X-ray irradiation and
the time of the $M$th fracture is
\begin{equation}
t_{tot}^M \simeq t_{break}^0
           \left[ 1 + (1 - k^{-1/3})
            \sum_{m=1}^M k^{-m(1 + 2\alpha/3)/3}\right] .
\end{equation}
This is a convergent series.  In the limit $M \gg 1$, the result approaches
\begin{equation}
t_{tot}^\infty \simeq t_{break}^0
            \left[ 1 + {1 - k^{-1/3} \over k^{(1 + 2\alpha/3)/3} - 1}\right]~,
\end{equation}
which exceeds $t_{break}^0$ by a modest factor.
For example, when $\alpha = 0$, the total time is simply $1 + k^{-1/3}$
times the time required for the first break.

     Although the approximation $x_{min} > x_K$ is broadly valid, it is not
universally applicable.  In carbon-rich grains, for which $x_K = 0.28$, it
is a valid approximation for fragments as small as $\simeq 18$~\AA.
Grains this small are hardly more than large molecules, so the
expressions just derived apply well to the entire destruction process for
carbon-rich grains.

    However, in silicate grains, for which $\langle x_K^3 \rangle \simeq 1$,
the approximation $x_{min} > x_K$ is only appropriate for
$a \ga 0.013 \, \mu m = 130$~\AA.
For smaller grains, $x_K > x_{min}$, and the nominal breaking time then 
becomes
\begin{equation}
t_{break} = D x_K^{\alpha} a_{0.1}^{-1} ;
\end{equation}
that is, the breaking time actually {\it increases} as the grain becomes
smaller.  This is because the charging rate is proportional to the volume
of the grain ($\propto a^3$), whereas the critical charge is $\propto a^2$.
Consequently, the total time to break silicate grains is however long it
takes to reach the $x_{min} < x_K$ regime plus an additional time
\begin{equation}
t_{*} \simeq  D x_{K}^{\alpha} a_{o,0.1}^{-1} (1 - k^{-1/3})
            \sum_{m=m_*+1}^M k^{m/3} ,
\label{tstar}
\end{equation}
where $a_{o,0.1}$ is the original grain size in units of $0.1 \micron$
and $m_*$ is the number of breaks necessary to reduce the
grain to a size such that
$x_{min} < x_K$.  The sum in Equation~\ref{tstar} is
$(k^{M/3} - k^{m_*/3})/(1 - k^{-1/3})$; thefore, 
\begin{equation}
t_{*} \simeq D x_{K}^{\alpha} a_{o,0.1}^{-1}\left(k^{M/3} - k^{m_*/3}\right).
\end{equation}
Here the two critical numbers of events are
\begin{equation}
m_* = 3\log_k \left(7.7 a_{o,0.1} x_{K}^{-3/2}\right) 
\end{equation}
and
\begin{equation}
M \simeq 3\log_k \left(33 a_{o,0.1}\right),
\end{equation}
where we have, in the second equation, assumed that complete breakdown of
a grain occurs when it has been reduced to 30~\AA\ in size.  For example, if
we take $k=3$, $m_* \simeq 6 + \log_3 (a_{o,0.1}/x_{K}^{3/2})$ and
$M \simeq 10 + \log_3 a_{o,0.1}$.  The factor $k^{M/3} - k^{m_*/3}$ is
then $(33 - 7.7 x_K^{-3/2})a_{o,0.1}$.  Although this is a relatively
large number, the absence of the
$4^{3+\alpha}$ factor from $t_*$ (as contrasted with $t_{tot}$) means that
the time spent in the $x_{min} < x_K$ regime can at most be comparable
to the time required for the initial break of a grain with $a \sim 0.1 \, \mu 
m$.

\begin{deluxetable}{llc}
\tablewidth{0pt}
\tablecaption{Glossary of Symbols}
\tablehead{\colhead{Symbol} & \colhead{Definition} & \colhead{Section}}
\startdata
$A$ & modified fluence:  $\sigma_{-19} n_{23} E_{51} D_{100}^{-2}$ & 3.2 \\
$a_{0.1}$ & grain size in units of $0.1$~$\mu$m & 3.2 \\
$B_\lambda(T)$ & Planck black body radiation function & 3.2 \\
$C$ & modified radiated X-ray energy: $=D_{100}^2A$ & 4.2\\
$D_{100}$& distance from burst to dust in units of 100 pc & 3.2 \\
$E_\epsilon$ & energy density of radiated X-rays in ergs keV$^{-1}$ & 3.2 \\
$E_{51}$ & fiducial energy radiated in X-rays:
$4\pi \epsilon dE_\epsilon/ d\Omega$ evaluated at $\epsilon = 1$~keV & 3.2\\ 
 & in units of $10^{51}$~erg & \\ 
$e$ & electron charge, with a sign convention that $e = -|e|$ & 3.3 \\
$G$ & rate of heating of a grain in erg s$^{-1}$ & 3.2\\
$H_{-11}$ & binding energy per atom in a grain units of $10^{-11}$ ergs & 3.2 
\\
$k_b$ & Boltzmann constant & 3.2 \\
$N$ & the H atom column density along the line of sight & 3.2 \\
$n_{23}$ & atomic density in a grain in units of $10^{23}$~cm$^{-3}$ & 3.2\\
$p$ & power-law index of the burst injected electron energy spectrum & 2 \\
${\cal Q}$ & total thermal energy in the grain in ergs & 3.2 \\
$Q_a$ & optical absorption efficiency of a spherical grain & 3.2 
\\
$\langle Q_A \rangle_T$ & absorption efficiency $Q_a$ averaged over a Planck
spectrum of temperature $T$ & 
3.2 \\
$R_{ion}$ & ionization rate of a grain in electrons per second & 3.3 \\
$S_{10}$ & stress across a grain in units of $10^{10}$ dyne cm$^{-2}$& 3.3 \\
$S_{crit}$ & stress sufficient to shatter a grain & 4.3 \\
$S_{crit,10}$ & $S_{crit}$ in units of $10^{10}$~dyne~cm$^{-2}$ & 4.3 \\
$T_3$ & temperature in units of $10^3$ K & 3.2 \\
$T_{e4} $ & electron temperature in units of $10^4$ K & 3.3 \\
$T_{eq} $ & equilibrium temperature of a grain & 3.2 \\
$T_{r=s}$ & temperature where radiation cooling equals sublimation cooling & 
3.2 \\
$T_{sub}$ & temperature required to sublimate a grain during the burst & 3.2 
\\
${\cal T}_h$ & flux weighted transparency to X-rays of the ISM & 3.2 \\
${\cal T}_i$ & photon-number weighted transparency to X-rays of the ISM & 3.3 
\\
$t_{10}$ & characteristic time of the X-ray emission scaled to 10~s & 3.2\\
\tablebreak
$x$ &  photon energy in keV & 3.2\\
$x_c$ & energy above which the ISM is effectively transparent to X-rays & 3.3 
\\
$x_K$ & energy in keV of the most important K-edge in the grain & 3.2 \\
$x_{min}$ & minimum energy (in keV) of electrons able to escape from grain &
3.2 \\
$Z$ & atomic number & 2 \\
$Z_g$ & grain charge in units of electron charge & 3.3 \\
$\alpha$ & energy spectral index of the burst X-ray flux & 2 \\
$\epsilon$ & photon energy & 2 \\
$\phi$ & a correction factor of order unity used in radiative cooling
equations& 3.2 \\
$\rho$ & density of a grain in gm cm$^{-3}$ & 3.1 \\
$\sigma_{-19}$ & K-shell edge cross section in units of $10^{-19}$~cm$^2$ &
3.2\\
\enddata
\tablecomments{Some symbols used in a single equation or paragraph and 
those used only in the appendix have been excluded.}
\end{deluxetable}

\end{document}